\documentclass[preprint,12pt]{elsarticle}
\usepackage{todonotes}
\usepackage{booktabs}
\usepackage{multirow}
\usepackage{colortbl}
\usepackage{subcaption}  
\usepackage{hyperref}
\usepackage{amsmath}
\usepackage{cleveref}

\usepackage[ruled,vlined]{algorithm2e}

\journal{Journal of Software and Systems}

\begin{document}
\begin{frontmatter}

\title{A Case Study on the Stability of Performance Tests for Serverless Applications}

\author[1]{Simon Eismann}
\ead{simon.eismann@uni-wuerzburg.de}

\author[2]{Diego Elias Costa}
\ead{diego.costa@concordia.ca}

\author[2]{Lizhi Liao}
\ead{l_lizhi@encs.concordia.ca}

\author[3]{Cor-Paul Bezemer}
\ead{bezemer@ualberta.ca}

\author[2]{Weiyi Shang}
\ead{shang@encs.concordia.ca}

\author[4]{Andr\'e van Hoorn}
\ead{van.hoorn@iste.uni-stuttgart.de}

\author[1]{Samuel Kounev}
\ead{samuel.kounev@uni-wuerzburg.de}

\address[1]{University of W\"urzburg, W\"urzburg, Germany}
\address[2]{Concordia University, Montreal, Canada}
\address[3]{University of Alberta, Edmonton, Canada}
\address[4]{University of Stuttgart, Stuttgart, Germany}

\begin{abstract}
\textit{Context.} While in serverless computing, application resource management and operational concerns are generally delegated to the cloud provider, ensuring that serverless applications meet their performance requirements is still a responsibility of the developers. Performance testing is a commonly used performance assessment practice; however, it traditionally requires visibility of the resource environment.

\textit{Objective.} In this study, we investigate whether performance tests of serverless applications are stable, that is, if their results are reproducible, and what implications the serverless paradigm has for performance tests.

\textit{Method.} We conduct a case study where we collect two datasets of performance test results: (a) repetitions of performance tests for varying memory size and load intensities and (b) three repetitions of the same performance test every day for ten months.

\textit{Results.} We find that performance tests of serverless applications are comparatively stable if conducted on the same day. However, we also observe short-term performance variations and frequent long-term performance changes.

\textit{Conclusion.} Performance tests for serverless applications can be stable; however, the serverless model impacts the planning, execution, and analysis of performance tests.

\end{abstract}

\end{frontmatter}

\section{Introduction}
Serverless computing combines Function-as-a-Service (e.g., AWS Lambda, Google Cloud Functions, or Azure Functions) and Backend-as-a-Service (e.g., managed storage, databases, pub/sub, queueing, streaming, or workflows) offerings that taken together provide a high-level application programming model, offloading application resource management and operation aspects to the cloud provider~\cite{kounev_et_al:DagRep.11.5.1:Ch5.1:ServerlessNotion, eismann2020review}. 
The cloud provider opaquely handles resource management tasks, such as deployment, resource allocation, or auto-scaling, and bills the user on a pay-per-use basis~\cite{baldini2017serverless, eyk2019Reference}.
While the cloud provider takes care of resource management, managing the performance of serverless applications remains a developer concern~\cite{leitner2019mixed, Eyk2018Challenges}.
Executing performance tests as part of a CI/CD pipeline to monitor the impact of code changes on system performance is a common and powerful approach to manage system performance~\cite{daly2020Change, leitner2017Exploratory}. One of the key requirements for reliable performance tests is ensuring that an identical resource environment is used for all tests~\cite{Eismann2020microservices}. 

However, with serverless applications, developers have no control over the resource environment. Worse yet, cloud providers expose no information to developers about the resource environment~\cite{wang2018peeking}. Therefore, information such as the number of provisioned workers, worker utilization, worker version, virtualization stack, or underlying hardware is unavailable. Furthermore, cold starts (requests where a new worker has to be provisioned) are a widely discussed performance challenge~\cite{leitner2019mixed, Eyk2018Challenges}. This begs the following question: \\
\\
``\emph{Are performance tests of serverless applications stable?}''\\

The performance variability of virtual machines in cloud environments has been studied extensively~\cite{Iosup2011on, 10.14778/1920841.1920902, laaber:19-emse}. However, many serverless platforms are not deployed on traditional virtual machines~\cite{agache2020firecracker}. Additionally, the opaque nature of serverless platforms means that it is extremely challenging, if not impossible, to control or know how many resources are allocated during a performance test~\cite{eyk2019Reference}. 
Existing work on performance evaluation of serverless platforms focuses on determining the performance characteristics of such platforms but does not investigate the stability of performance measurements~\cite{wang2018peeking, 7979855, lloyd2018serverless, scheuner:20-jss}. 

In this paper, we present an exploratory case study on the stability of performance tests of serverless applications. Using the serverless airline application~\cite{buildon}, a representative, production-grade serverless application~\cite{reinvent}, we conduct two sets of performance measurements: (1) multiple repetitions of performance tests under varying configurations to investigate the performance impact of the latter, and (2) three daily measurements for ten months to create a longitudinal dataset and investigate the stability of performance tests over time.

Based on these sets of measurements, we investigate the following research questions:
\begin{itemize}
    \item \textbf{RQ1:} How do cold starts influence the warm-up period and stability of serverless performance tests?
    \item \textbf{RQ2:} How stable are the performance test results of a serverless application deployed on public serverless platforms?
    \item \textbf{RQ3:} Does the performance of serverless applications change over time?
\end{itemize}

We find that there are serverless-specific changes and pitfalls to all performance test phases: design, execution, and analysis. In the design phase, the load intensity of the test directly correlates to cost, and reducing load intensity can deteriorate performance. In the execution phase, daily performance fluctuations and long-term performance changes impact the decision when performance tests should be scheduled. In the analysis phase, developers need to consider that there is still a warm-up period after removing all cold starts and that cold starts can occur late in a performance test even under constant load. Our reported findings in this paper can be used as guidelines to assist researchers and practitioners in conducting performance tests for serverless applications.

The rest of the paper is organized as follows: Section~\ref{sec:background} gives an introduction to performance testing and serverless applications. Next, Section~\ref{sec:related} introduces related work on performance evaluation of serverless platforms as well as on performance variability of virtual machines. Section~\ref{sec:study_design} describes the design of our case study and Section~\ref{sec:results} analyses the results in the context of the three research questions. Then, Section~\ref{sec:discussion} discusses the implications of our findings on the performance testing of serverless applications. Section~\ref{sec:validity} presents the threats to the validity of our study, Section~\ref{replication} introduces our comprehensive replication package, and finally, Section~\ref{sec:conclusion} concludes the paper.

\section{Background}
\label{sec:background}
In the following, we give a short introduction to serverless applications and performance testing.

\subsection{Serverless Applications}
Serverless applications consist of business logic in the form of serverless functions---also known as Function-as-a-Service (FaaS)---and cloud provider-managed services such as databases, blob storage, queues, pub/sub messaging, machine learning, API gateways, or event streams. 

Developers write business logic as isolated functions, configure the managed services via Infrastructure-as-Code, and define triggers for the execution of the business logic (serverless functions). Triggers can either be HTTP requests or cloud events such as a new message in a queue, a new database entry, a file upload, or an event on an event bus. The developer provides the code for the serverless functions, the configuration of the managed services, and the triggers; the cloud provider guarantees that the code is executed and the managed services are available whenever a trigger occurs, independent of the number of parallel executions. In contrast to classical IaaS platforms, developers are not billed for the time resources are \emph{allocated} but rather for the time resources are actively \emph{used}. Under this pay-per-use model, developers are billed based on the time the serverless functions run and per operation performed by the managed services.

Serverless functions promise seamless scaling of arbitrary code. In order to do so, each function is ephemeral, stateless, and is executed in a predefined runtime environment. Additionally, every worker (function instance) only handles a single request at a time.
When a function trigger occurs, the request is routed to an available function instance. If no function instance is available, the request is not queued, instead, a new function instance is deployed and the request is routed to it (known as \emph{cold start}, other executions are labeled as \emph{warm start}). In order to keep the time for a cold start in the range of a few hundred milliseconds, the platform utilizes a fleet of template function instances that already run the supported runtime environments, into which only the application-specific code needs to be loaded.

A number of potential benefits of serverless applications compared to traditional cloud applications have been reported~\cite{eismann2020serverless}. The pay-per-use model is reported to reduce costs for bursty workloads, which often lead to overprovisioning and low resource utilization in traditional cloud-based systems. Furthermore, serverless applications are virtually infinitely scalable by design and reduce the operational overhead, as the cloud provider takes care of all resource management tasks. Finally, the heavy usage of managed services is reported to increase development speed.

\subsection{Performance Testing}

Performance testing is the process of measuring and ascertaining a system's performance-related aspects (e.g., response time, resource utilization, and throughput) under a particular workload~\cite{DBLP:journals/tse/JiangH15}. Performance testing helps to determine compliance with performance goals and requirements~\cite{DBLP:journals/pcs/PozinG11, DBLP:journals/tse/WeyukerV00, DBLP:journals/iee/KalitaB11}, identify bottlenecks in a system~\cite{DBLP:conf/icsm/SyerJNHNF13, DBLP:conf/msr/NguyenNHNF14}, and detect performance regressions~\cite{DBLP:conf/wosp/NguyenAJHNF12, DBLP:conf/icse/MalikHH13, DBLP:conf/wosp/ShangHNF15}.
A typical performance testing process starts with designing the performance tests according to the performance requirements. These performance tests are then executed in a dedicated performance testing environment, while the system under test (SUT) is continuously monitored to collect system runtime information including performance counters (e.g., response time and CPU utilization), the system's execution logs, and event traces. Finally, performance analysts analyze the results of the performance testing.

During the execution of a software system, it often takes some time to reach its stable performance level under given load. 
During performance testing, the period before the software system reaches steady-state is commonly known as the warm-up period, and the period after that is considered as the steady-state period.
There are many reasons for the warm-up period, such as filling up buffers or caches, program JIT compilation, and absorbing temporary fluctuations in system state~\cite{DBLP:conf/compsac/MansharamaniKMS10}.
Since performance during the warm-up period may fluctuate, in practice, performance engineers often remove the duration of the unstable phase (i.e., warm-up period) of the performance test and only consider the steady-state period in the performance test results.
The most intuitive way to determine the warm-up period is to simply remove a fixed duration of time (e.g., 30~minutes~\cite{DBLP:journals/ese/LiaoCLZSGSTS20}) from the beginning of the performance testing results.
We refer to a review by Mahajan and Ingalls~\cite{DBLP:conf/wsc/MahajanI04} for an overview of existing techniques to determine the warm-up period.

\section{Related Work}
\label{sec:related}
Existing work related to this study can be grouped into performance evaluations of serverless platforms and studies on the performance variability of virtual machines.

\subsection{Performance Evaluation of Serverless Platforms}

A number of empirical measurement studies on the performance of serverless applications have been conducted. Lloyd et al.~\cite{lloyd2018serverless} examined the infrastructure elasticity, load balancing, provisioning variation, infrastructure retention, and memory reservation size of AWS Lambda and Azure Functions. They found that cold and warm execution times are correlated with the number of containers per host, which makes the number of containers per host a major source of performance variability. 
Wang et al. conducted a large measurement study that focuses on reverse engineering platform details~\cite{wang2018peeking}. They found variation in the underlying CPU model used and low performance isolation between multiple functions on the same host. When they repeated a subset of their measurements about half a year later, they found significant changes in the platform behavior.
Lee et al.~\cite{Lee2018} analyzed the performance of CPU, memory, and disk-intensive functions with different invocation patterns. They found that file I/O decreases with increasing numbers of concurrent requests and that the response time distribution remained stable for a varying workload on AWS. %
Yu et al.~\cite{yu2020characterizing} compared the performance of AWS Lambda to two open-source platforms, OpenWhisk and Fn. They found that Linux CPU shares offer insufficient performance isolation and that performance degrades when co-locating different applications. %
However, while the performance of FaaS platforms has been extensively studied, there has been little focus on the stability of these measurements over time.

There have also been a number of measurement tools and benchmarks developed for serverless applications and platforms. Cordingly et al.~\cite{cordingly2020predicting} introduced the Serverless Application Analytics Framework (SAAF), a tool that allows profiling FaaS workload performance and resource utilization on public clouds; however it does not provide any example applications. 
Figiela et al.~\cite{figiela2018performance} introduced a benchmarking suite for serverless platforms and evaluated the performance of AWS Lambda, Azure Functions, Google Cloud Functions, and IBM Functions. The workloads included in the benchmarking suite consist of synthetic benchmark functions, such as a mersenne
 twister or linpack implementation. 
Kim et al.~\cite{8814583} proposed FunctionBench, a suite of function workloads for various cloud providers. The functions included in FunctionBench closely resemble realistic workloads, such as video processing or model serving, but they only cover single functions and not entire applications.
In summary, there has been a strong focus on benchmarks and tooling around FaaS but much less focus on realistic applications.

For further details on the current state of the performance evaluation of serverless offerings, we refer to an extensive multi-vocal literature review by Scheuner et al.~\cite{scheuner:20-jss}. This review also finds that the reproducibility of the surveyed studies is a major challenge. %

\subsection{Performance Variability of Virtual Machines}
Due to the extensive adoption of virtual machines (VMs) in practice, there exists much prior research on the performance variability of VMs.
One early work in this area is from Menon et al.~\cite{DBLP:conf/vee/MenonSTJZ05}, which quantifies the network I/O related performance variation in Xen virtual machines. Their results also identify some key sources of such performance variability in VMs. 
Afterwards, Kraft et al.~\cite{DBLP:conf/wosp/KraftCKGK11}, Boutcher and Chandra~\cite{DBLP:journals/sigops/BoutcherC10} apply various techniques to assess the performance variation of VMs compared to a native system with respect to disk I/O.
Taking the contention between different VMs into account, Koh et al.~\cite{DBLP:conf/ispass/KohKBBWP07} analyze ten different system-level performance characteristics to study the performance interference effects in virtual environments. Their results show that the contention on shared physical resources brought by virtualization technology is one of the major causes of the performance variability in VMs.

Huber et al.~\cite{DBLP:conf/closer/HuberQHK11} compared the performance variability (for CPU and memory) of two virtualization environments and use regression-based models to predict the performance overhead for executing services on these platforms.
Schad et al.~\cite{10.14778/1920841.1920902}, Iosup et al.~\cite{Iosup2011on}, and Leitner and Cito~\cite{Leitner2016patterns} assessed the performance variability across multiple regions and instance types of popular public clouds such as Amazon Web Services (AWS) and Google Compute Engine (GCE).
Based on these findings, Asyabi~\cite{DBLP:journals/fgcs/AsyabiSB18} proposed a novel hypervisor CPU scheduler aiming to reduce the performance variability in virtualized cloud environments.

To investigate the impact of the performance variability of VMs on performance assurance activities (e.g., performance testing and microbenchmarking), Laaber et al.~\cite{DBLP:conf/msr/LaaberL18, laaber:19-emse} evaluated the variability of microbenchmarking results in different virtualization environments and analyzed the results from a statistical perspective. They found that not all cloud providers and instance types are equally suited for performance microbenchmarking. 
Costa et al.~\cite{8747433} summarized some bad practices of writing microbenchmarks using the JMH framework to mitigate the variation and instability of cloud environments when conducting performance microbenchmarking.
Arif et al.~\cite{DBLP:journals/ese/ArifSS18} and Netto et al.~\cite{DBLP:conf/ipps/NettoMVCOSZ11} compared performance metrics generated via performance tests between virtual and physical environments. Their findings highlight the inconsistency between performance testing results in virtual and physical environments.

Prior research has extensively studied the performance variability of VMs, also proposing approaches to mitigate it.
However, the focus of existing work is on traditional software systems in virtualized environments with pre-known configurations.
In comparison, in this work, we consider studying the stability of performance tests for serverless applications where developers have no information about (and control of) the virtualization resource environment.

\begin{figure*}[t]
    \centering
    \includegraphics[width=\textwidth]{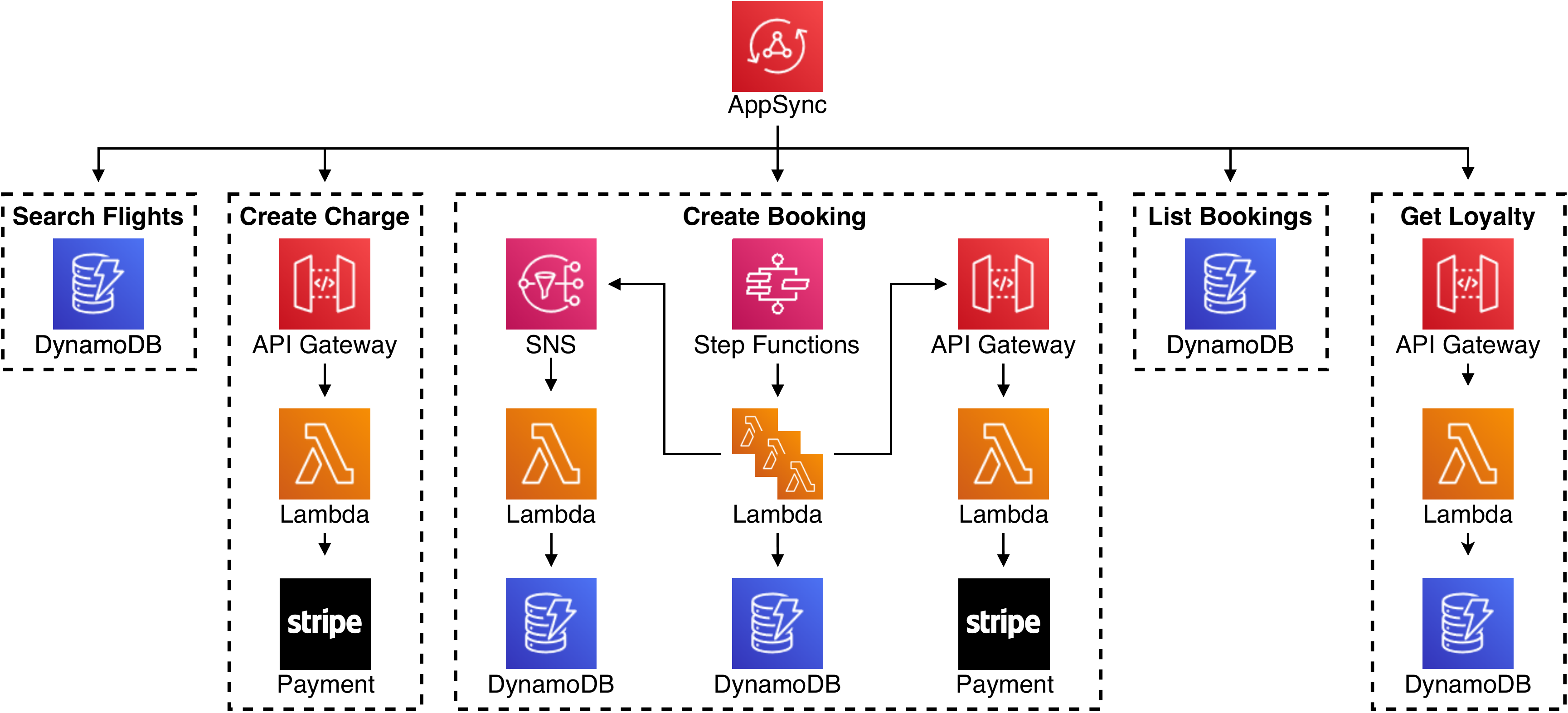}
    \caption{Architecture and API endpoints of the serverless airline booking application.}
    \label{fig:airline_architecture}
\end{figure*}

\section{Case Study Design}
\label{sec:study_design}

In this section, we present the design of our case study. We first introduce our subject system, the Serverless Airline Booking  (SAB) application, and describe why we selected this system for our case study. Then, we describe the experiment setup, the collected metrics, and the individual experiments. 

\subsection{Serverless Airline Booking (SAB)}
\label{sec:airline}
The serverless airline booking application (SAB)\footnote{\url{https://github.com/aws-samples/aws-serverless-airline-booking}} is a fully serverless web application that implements the flight booking aspect of an airline on AWS. It was presented at AWS re:Invent as an example for the implementation of a production-grade full-stack app using AWS Amplify~\cite{reinvent}. The SAB was also the subject of the AWS Build On Serverless series~\cite{buildon}. Customers can search for flights, book flights, pay using a credit card, and earn loyalty points with each booking. 

The frontend of the SAB is implemented using CloudFront, Amplify/S3, Vue.js, the Quasar framework, and Stripe Elements. 
This frontend sends GraphQL queries (resolved using AWS AppSync) to five backend APIs, as shown in Figure~\ref{fig:airline_architecture}: 
\begin{itemize}
    \item The \emph{Search Flights} API retrieves all flights for a given date, arrival airport, and departure airport from a DynamoDB table using the DynamoDB GraphQL resolver.
    \item The \emph{Create Charge} API is implemented as an API gateway that triggers the execution of the \emph{CreateStripeCharge} lambda function, which manages the call to the Stripe API. 
    \item The \emph{Create Booking} API reserves a seat on a flight, creates an unconfirmed booking, and attempts to collect the charge on the customer's credit card. If successful, it confirms the booking, and awards loyalty points to the customer. In case the payment collection fails, the reserved seat is freed again, and the booking is canceled. This workflow is implemented as an AWS Step Functions workflow that coordinates multiple lambda functions. The functions \emph{ReserveBooking} and \emph{CancelBooking} directly modify DynamoDB tables, the \emph{NotifyBooking} function publishes a message to SNS, which is later consumed by the \emph{IngestLoyalty} function that updates the loyalty points in a DynamoDB table. The \emph{CollectPayment} and \emph{RefundPayment} functions call the Stripe backend via an application from the Serverless Application Repository.
    \item The \emph{List Bookings} API retrieves the existing bookings for a customer. Similar to the \emph{Search Flights} API, this is implemented using a DynamoDB table and the DynamoDB GraphQL resolver. 
    \item The \emph{Get Loyalty} API retrieves the loyalty level and loyalty points for a customer. An API Gateway triggers the lambda function \textit{FetchLoyalty}, which retrieves the loyalty status for a customer from a DynamoDB table.
\end{itemize}
We selected SAB for our case study after investigating potential applications from a review of serverless use cases~\cite{eismann2020review}, a serverless literature dataset~\cite{Al-Ameen2018Systematic}, and a recent survey on FaaS performance evaluation~\cite{scheuner:20-jss}. We chose SAB over other potential applications due to its comparatively large size and its usage of many different managed services. It is also running on AWS, the by far most popular cloud provider for serverless applications~\cite{eismann2020review,leitner2019mixed,nuweba}, and it uses both Python and JavaScript to implement the serverless functions, the two most popular programming languages for serverless applications~\cite{eismann2020review,leitner2019mixed,nuweba}.

\subsection{Experiment Setup}
\label{sec:setup}
We deploy the frontend via Amplify~\cite{amplify} and the backend services via either the Serverless Application Model~\cite{sam} or CloudFormation~\cite{cloudformation} templates depending on the service. The serverless nature of the application makes it impossible to specify the versions of any of the used services, as DynamoDB, Lambda, API Gateway, Simple Notification Service, Step Functions, and AppSync all do not provide any publicly available version numbers.

For the load profile, customers start by querying the \textit{Search Flights} API for flights between two airports. If no flight exists for the specified airports and date, the customer queries the \textit{Search Flights} API again, looking for a different flight. We populated the database so that most customers find a flight within their first query. Next, they call the \textit{Create Charge} API and the \textit{Create Booking} to book a flight and pay for it. After booking a flight, each customer checks their existing bookings and loyalty status via the \textit{List Bookings} API and the \textit{Get Loyalty} API. This load profile is implemented using the TeaStore load driver~\cite{kistowski2018TeaStore}. 

In terms of monitoring data, we collect the response time of each API call via the load driver. Additionally, we collect the duration, that is, the execution time of every lambda function. We exclude the duration of the lambdas \textit{ChargeCard} and \textit{FetchLoyalty}, as the response times of the APIs \textit{Create Charge} and \textit{Get Loyalty} mostly consist of the execution times of these lambdas. We cannot collect any resource-level metrics such as utilization or number of provisioned workers, as AWS and most other major serverless platforms do not report any resource level metrics.

For our experiments, we perform measurements with 5 req/s, 25 req/s, 50 req/s, 100 req/s, 250 req/s, and 500 req/s to cover a broad range of load levels. Additionally, we vary the memory size of the lambda functions between 256\,MB, 512\,MB,  and 1024\,MB, which covers the most commonly used memory sizes~\cite{datadog}. For each measurement, the SAB is deployed, put under load for 15 minutes, and then torn down again. We perform ten repetitions of each measurement to account for cloud performance variability. Additionally, we run the experiments as randomized multiple interleaved trials, which have been shown to further reduce the impact of cloud performance variability~\cite{Abedi2017conducting}. To minimize the risk of manual errors, we fully automate the experiments (for further details see Section~\ref{replication}). These measurements started on July 5th, 2020, and continuously ran until July 17th, 2020.

Additionally, we set up a longitudinal study that ran three measurement repetitions with 100 req/s and 512 MB every day at 19:00 from \mbox{Aug 20th, 2020} to \mbox{Jun 20th, 2021}. The measurements were automated by a Step Functions workflow that is triggered daily by a CloudWatch alarm and starts the experiment controller VM, triggers the experiment, uploads the results to an S3 bucket, and shuts down the experiment controller VM again.

To ensure reproducibility of our results, the fully automated measurement harness and all collected data from these experiments are available in our replication package.\footnote{\url{https://github.com/ServerlessLoadTesting/ReplicationPackage}}

\section{Case Study Results}
\label{sec:results}
We now present the results of our empirical study in the context of our three research questions. For each research question, we present the motivation of answering the question, our approach to answering it, and the corresponding results. 

\subsection{RQ1: How do cold starts influence the warm-up period and stability of serverless performance tests?}
\label{sec:rq1}

\begin{algorithm}[tb]
\SetAlgoLined
\KwResult{warmupInSeconds}
 threshold = 0.01\;
 stable = False\;
 warmupInSeconds = 0\;
 global\_mean = mean(ts)\;
 \While{stable == False}{
  ts = remove5secs(ts) // Remove 5 seconds of data\;
  warmupInSeconds += 5\;
  new\_mean = mean(ts)\;
  delta = abs((new\_mean - global\_mean) / global\_mean)\;
  \eIf{delta $<$ threshold}{
    stable = True\;
  } {
    global\_mean = new\_mean\;
  } 
 }
 \caption{Warm-up Period Identification Heuristic.}
 \label{fig:warmup-heuristic}
\end{algorithm}

\noindent
\textbf{Motivation.}
A common goal of performance tests is to measure the steady-state performance of a system under a given workload. 
Hence, it is essential that practitioners understand how long it takes for serverless applications to reach stable performance (i.e., how long is the warm-up period) in order to plan the duration of their performance tests accordingly. 
Aside from the general aspects that influence the initial performance instability, such as the environment and application optimizations (e.g., CPU adaptive clocking and cache setup), serverless applications also encounter cold starts.
A cold start occurs when a request cannot be fulfilled by the available function instances, and a new instance has to be provisioned to process the upcoming request. Cold starts can incur significantly higher response times~\cite{wang2018peeking, figiela2018performance}.
Hence, in this RQ, we investigate: (1) how long is the warm-up period in our experiments and (2) the role of cold starts in the stability of the warm-up and steady-state experiment phases. 
\\

\noindent
\textbf{Approach.}
To determine the duration of the warm-up period, we initially tried to use the MSER-5 method~\cite{White:00:MSER5}, which is the most popular method to identify the warm-up period in simulations~\cite{DBLP:conf/wsc/MahajanI04, hoad2010automating}. However, this approach was not applicable due to the large outliers present in our data, a well-documented flaw of MSER-5~\cite{sandikcci2006analysis}. Therefore, we employ a heuristic to identify the warm-up period. 
Our heuristic, shown in Algorithm~\ref{fig:warmup-heuristic}, gradually removes data from the beginning of the experiment in windows of five seconds and evaluates the impact of doing so on the overall mean results.
If the impact is above a threshold (we used 1\% in our experiments), we continue the data removal procedure. 
Otherwise, we consider the seconds removed as the warm-up period and the remainder as the steady-state phase of the performance test experiment.

To evaluate the impact of cold starts on the experiment stability, we analyze the distribution of cold start requests across the two phases of performance tests: warm-up period and steady-state period. 
Then, we evaluate the influence of cold start requests on the overall mean response time, considering only cold start requests that occurred after the warm-up period. 
To test for statistically significant differences, we use the unpaired and non-parametric Mann-Whitney U test~\cite{mann1947}. 
In cases where we observe a statistical difference, we evaluate the effect size of the difference using the Cliff's Delta effect size~\cite{Cliff}, and we use the following common thresholds~\cite{Romano} for interpreting the effect size:

\begin{equation*}
	\text{Effect size \emph{d}}=\begin{cases}
	negligible (N), & \text{if $|d| \leq 0.147$}\\
	small (S), & \text{if  $0.147 <|d| \leq 0.33$}\\
	medium (M), &\text{if $0.33 < |d| \leq 0.474$}\\
	large (L), &\text{if $0.474 < |d| \leq 1$}\\
	\end{cases}
\end{equation*}

Note that not all request classes provide information about cold starts. This information is only available for the six lambda functions, as the managed services either do not have cold starts or do not expose them.
Therefore, we report the cold start analysis for the following six request classes: \textit{CollectPayment}, \textit{ConfirmBooking}, \textit{CreateStripeCharge}, \textit{IngestLoyalty}, \textit{NotifyBooking}, and \textit{ReserveBooking}. 
Finally, our experiment contains more than 45 hours of measurements, including performance tests with ten repetitions, different workload levels, and function sizes. \\

\begin{table}[tb]
    \centering
    \caption{\textbf{Maximum} warm-up period in seconds across ten repetitions of all function sizes. We highlight warm-up periods over one minute with a dark background.}
    \label{tab:warmup-period}
    
\newcommand{\gc}{\cellcolor{gray!25}}
\newcommand{\dc}{\cellcolor{gray!50}}

\begin{tabular}{lrrrrrr}
\toprule
\multirow{2}{*}{\textbf{Request Class}}  & \multicolumn{6}{c}{\textbf{Workload (reqs/s)}} \\
 &    \textbf{5}   &   \textbf{25}  &   \textbf{50}  &   \textbf{100} &   \textbf{250} &    \textbf{500} \\
 
\midrule
CollectPayment     &   15 &   10 &   10 &   10 &   10 &   10 \\
ConfirmBooking     &   25 &   15 &   15 &   \gc65 &   10 &   15 \\
CreateStripeCharge &   15 &   15 &   15 &   15 &   20 &   15 \\
Get Loyalty        &   \gc70 &   \gc60 &   45 &   30 &   10 &  -- \\
IngestLoyalty      &   15 &   25 &   55 &   \gc75 &  \dc125 &  \dc155 \\
List Bookings      &  \gc115 &   \gc70 &   55 &   45 &   25 &   15 \\
NotifyBooking      &   20 &   40 &   40 &   15 &   \gc60 &   10 \\
Process Booking    &   \gc65 &   45 &   20 &   10 &   10 &   15 \\
ReserveBooking     &   20 &   20 &   10 &   10 &   10 &   15 \\
Search Flights     &  \dc135 &   \gc80 &   50 &   30 &   30 &   20 \\
\bottomrule

\end{tabular}
\end{table}

\begin{table}[tb]
    \centering
    \caption{Average occurrence of cold start requests in the performance tests per request class. We consider the first 2 minutes as the warm-up period. We report cold starts as impacting the results if there is a significant difference of the mean response time when accounting for cold start requests after the warm-up period.}
    \label{tab:cold-start}
    \begin{tabular}{lrrrr}
\toprule
\multirow{2}{*}{\textbf{Request Class}} 
   & \textbf{\% Cold}    & \multicolumn{2}{c}{\textbf{\% Occurrence}}   & \multirow{2}{*}{\textbf{Impact?}} \\
    & \textbf{Start}       & \textbf{$<=$2 min}  & \textbf{$>$2 min}  &  \\

\midrule
CollectPayment      &  0.93 & 99.5 & 0.05 & No \\
ConfirmBooking      &  0.72 & 99.5 & 0.05 & No \\
CreateStripeCharge  &  0.44 & 99.9 & 0.01 & No \\
IngestLoyalty       &  1.01 & 99.2 & 0.02 & No \\
NotifyBooking       &  1.04 & 99.9 & 0.01 & No\\
ReserveBooking      &  0.40 & 99.8 & 0.02 & No \\
\bottomrule

\end{tabular}

\end{table}

\noindent
\textbf{Findings.}
\textbf{The warm-up period lasts less than 2 minutes in the vast majority of our experiments.} \Cref{tab:warmup-period} shows the maximum warm-up period in seconds, observed across all experiments per workload level. 
In most experiments, we observe that the maximum warm-up period out of the ten repetitions lasts less than 30 seconds (37 out of 48 experiment combinations).
With exception of \textit{IngestLoyalty}, all workload classes exhibit a shorter warm-up period as the load increases. 
The average warm-up period in experiments with 500 requests per second was 27 seconds, half of the warm-up period observed in runs with 5 requests per second (52 seconds). 
The function \textit{Get Loyalty} never reaches a steady-state under high load, as it implements the performance anti-pattern ``Ramp'' due to a growing number of entries in the database~\cite{smith2002new}.
We also note that, contrary to the workload, the function size (memory size) has no influence on the warm-up period: in most cases, the difference of the warm-up period across function sizes (256~MB, 512~MB, 1024~MB) is not significant ($p>0.05$), with a negligible effect size for the few significantly different cases ($d < 0.147$).
In the following, we opt to conservatively consider the first 2 minutes of performance tests as part of the warm-up period for any subsequent analysis.

\textbf{The vast majority ($>$99\%) of cold starts occur during the first two minutes of the performance test (warm-up period). Cold start requests that occur after the warm-up period ($<$1\%) do not impact the measurements.}
\Cref{tab:cold-start} depicts the average percentage of cold start requests across different request classes, the share of cold start requests that occur in the warm-up period, and whether cold starts after the warm-up period significantly impact the mean response time. 
We consider cold starts to impact the results, if there is a significant difference between the mean response time with and without cold starts in the steady-state experiment phase. 
As we observe similar results in all six request classes, below we discuss only the \textit{CollectPayment} results.
On average, cold start requests in \textit{CollectPayment} make up for 0.93\% of the total number of requests. 
However, since they mostly concentrate in the first two minutes of the experiment (99.5\%), they are discarded from the final results as part of the warm-up period.
The remaining cold start requests (0.5\%) that occur throughout the run of our performance test did not significantly impact the response time (Mann-Whitney U test with $p > 0.05$).

\textbf{In the majority of experiments, removing the cold starts does not shorten the warm-up period.} Given that cold starts occur mostly during the warm-up period, we wanted to assess if the warm-up period is composed solely of cold start requests. 
Is it enough to simply drop cold start requests from the experiment and consider all other requests as part of the steady-state performance measurements?
\Cref{tab:warmup-difference} shows the difference of the warm-up period considering all requests (the one shown in \Cref{tab:warmup-period}), versus the warm-up period calculated by filtering the cold start requests from the experiment. 
In the majority of the experiments (22 out of 36 combinations), we observe no difference between dropping or keeping the cold start requests in the duration of the warm-up period.
Some request classes, however, exhibited shorter periods of warm-up once we filter out cold start requests, as the high response time of cold start requests contributes to the warm-up period.
For instance, the experiment with \textit{CreateStripesCharge} showed a consistent reduction of the warm-up period of at least 5 seconds (our heuristic's window size) for all the workload sizes. 
It is important to note, however, that the warm-up period\,---\,while shorter in some classes\,---\,is not only influenced by cold starts.

\subsection{RQ2: How stable are the performance test results of a serverless ap-plication deployed on common serverless platforms?}
\label{sec:rq2}

\noindent
\textbf{Motivation.}
In RQ1, we found that within a run, the results of a performance test quickly become stable.
The period of instability (warm-up) usually lasts less than two minutes, and the number of cold start requests that occur after this period does not impact the performance test results. 
However, results across multiple runs are likely to vary considerably.
Practitioners have no way to ensure that two different performance tests are executed in similar resource environments, given that deployment details in serverless applications are hidden from developers.
Hence, for this RQ, we study how the inherent variance in deployed serverless applications impacts the stability \emph{between} performance tests.
\\

\begin{table}[t]
    \centering
    \caption{Difference of the maximum warmup-period in seconds between experiments including all requests vs.\  experiments filtering out the cold start requests.}
    \label{tab:warmup-difference}
    \begin{tabular}{lrrrrrr}
\toprule
\multirow{2}{*}{\textbf{Request Class}}  & \multicolumn{6}{c}{\textbf{Workload (reqs/s)}} \\
 &    \textbf{5}   &   \textbf{25}  &   \textbf{50}  &   \textbf{100} &   \textbf{250} &    \textbf{500} \\
\midrule
CollectPayment     &   -5 &   -- &   -5 &   -- &   -- &   -- \\
ConfirmBooking     &   -- &   -- &   -- &   -- &   -- &   -- \\
CreateStripeCharge &  -10 &   -5 &   -5 &   -5 &   -5 &  -10 \\
IngestLoyalty      &   -- &   -- &  -10 &   -- &   -- &  -10 \\
NotifyBooking      &   -- &  -20 &  -20 &   -- &   -- &   -- \\
ReserveBooking     &   -5 &   -5 &   -- &   -- &   -- &   -- \\
\bottomrule
\end{tabular}
\end{table}

\noindent
\textbf{Approach.}
In this analysis, we evaluate the variation of the mean response time across experiment runs and study the influence of experiment factors such as the load level and function size.
We focus on evaluating the steady-state performance of performance tests. Hence, we discarded the data from the first two minutes of the performance test runs (warm-up period) and calculated the mean response time for the steady-state phase, that is, the remaining 13 minutes of experiment data. 

To evaluate the stability of the mean response time across runs, we first exclude outliers within an experiment that fall above the .99 percentile.
Then, we calculate the coefficient of variation of the response time across the ten repetitions, per workload level and function size. 
The coefficient of variation is the ratio of the standard variation to the mean and is commonly used as a metric of relative variability in performance experiments~\cite{cordingly2020predicting,Leitner2016patterns}. 
Similarly to RQ1, we test statistically significant differences using the Mann-Whitney U test~\cite{mann1947} and assess the effect size of the difference using the Cliff's Delta effect size~\cite{Cliff}.
\\

\begin{figure*}[t]
    \centering
    \includegraphics[width=1.1\linewidth]{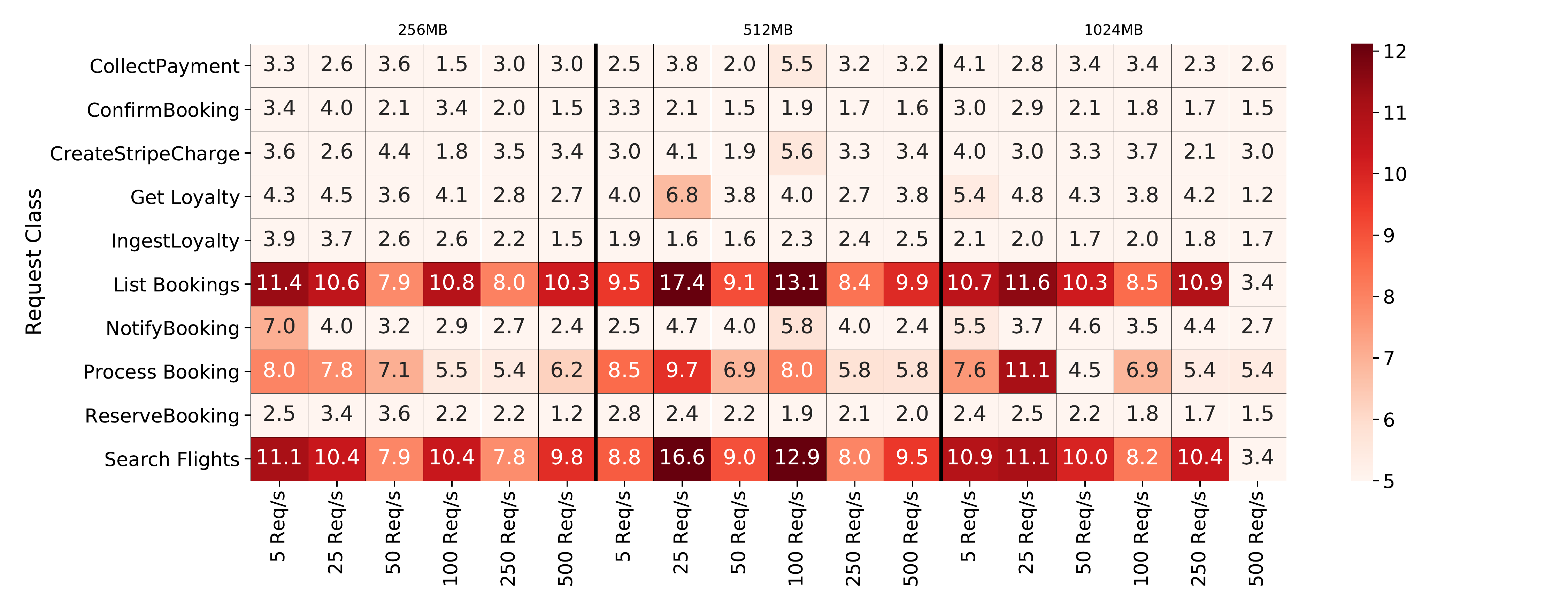}
    \caption{Coefficient of variation of the .99 mean across 10 repetitions per request class, load level, and function size. We highlight in the heatmap coefficients above 5\% of the mean.}
    \label{fig:coeff-variation}
\end{figure*}

\begin{figure}
    \centering
    \includegraphics[width=0.8\linewidth]{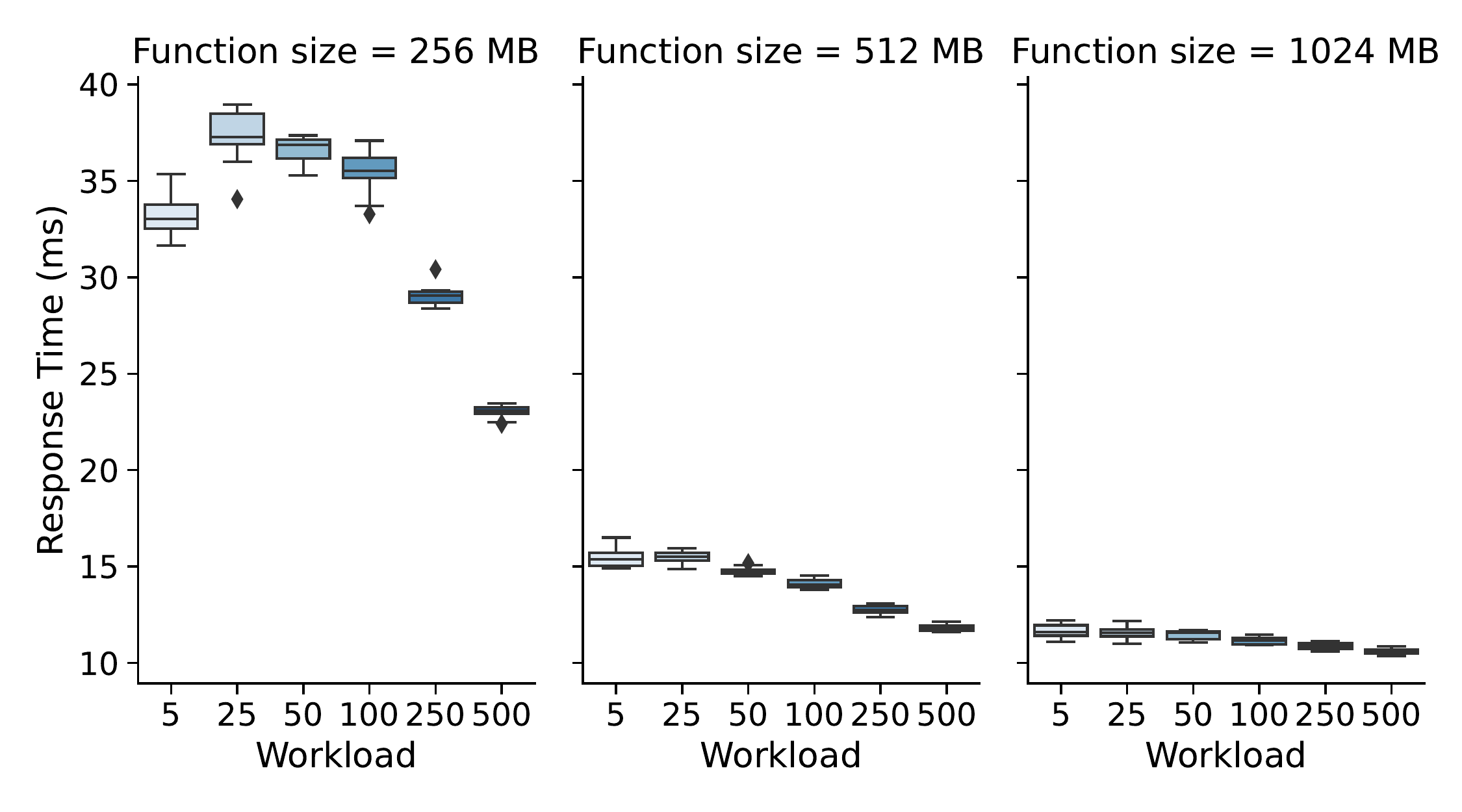}
    \caption{Response time of ten repetitions of \textit{ConfirmBooking} performance tests, per workload level and function size.}
    \label{fig:response-time-example}
\end{figure}

\noindent
\textbf{Findings.}
\textbf{We observe that the vast majority of experiments (160 out of 180) exhibits a coefficient of variation below 10\% of the mean response time.}
\Cref{fig:coeff-variation} shows a heat map of the coefficient of variation observed in 10 repetitions of all experiments.
With the exception of three request classes, \textit{List Booking}, \textit{Process Booking}, and \textit{Search Flights}, most of the other experiments show a coefficient of variation of less than 5\% of the mean (125 out of the 132 experiments).
The observed coefficient of variation is also in line with reported variation in other serverless benchmarks~\cite{cordingly2020predicting}, which was reported to be 5 to 10\% when executing synthetic workloads in the AWS infrastructure. 
This suggests that the studied serverless application performance tests are more stable than most traditional performance tests of cloud applications (IaaS).
Cito and Leitner~\cite{Leitner2016patterns} reported that performance variations of performance tests in cloud environments are consistently above 5\%, reaching variations above 80\% of the mean in several I/O-based workloads. 
The two classes with higher variability of the results, \textit{List Booking}, and \textit{Search Flights}, both use an Amplify resolver to retrieve data from DynamoDB without a lambda. Our findings indicate that this AWS-managed resolver might suffer from a larger performance variability.

\begin{figure}
    \centering
    \includegraphics[width=0.8\linewidth]{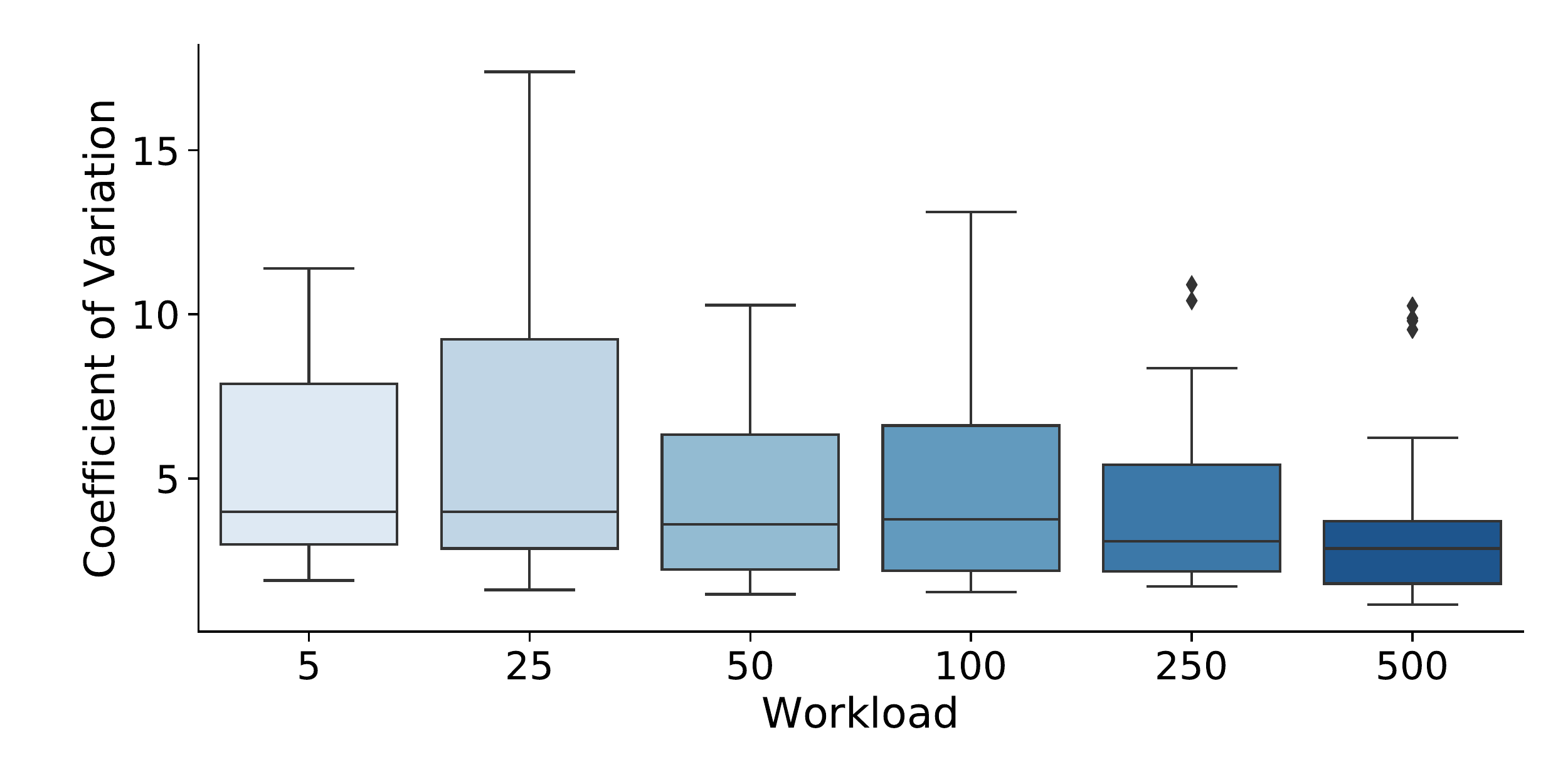}
    \caption{Distribution of coefficients of variation across all request classes and function size, per workload (reqs/s).}
    \label{fig:coefficient-per-workload}
\end{figure}

\textbf{We observe improvement in the response time and result stability in scenarios with higher workloads.}
\Cref{fig:response-time-example} shows the response time of the \textit{ConfirmBooking} request class, in which we observe that as the workload increases, the average response time decreases for all function sizes. 
This is true across almost all experiments, where the response time observed in the scenario with 500 requests per second is significantly faster than scenarios with only 5 requests per second (Mann-Whitney U $p < 0.05$), often to large effect sizes (Cliff's delta $d > 0.474$).
Moreover, the stability of the obtained average response time (across 10 repetitions) also improves slightly, from 4.6\% on average across all experiments with 5 reqs/s, to 3.3\% on experiments with 500 reqs/s (see \Cref{fig:coefficient-per-workload}).
Our findings suggest that the workload in the studied serverless application showed an inverse relationship to measured performance, that is, the higher the workload we tested the faster was the average response time, the opposite of what is expected in most typical systems (bare-metal, cloud environments).
It is important to note that, given the cost model of serverless infrastructure, performance tests with 500 requests per second cost 100~x more than tests with 5 requests per second.
Therefore, the small gain in stability is unlikely to justify the much higher costs of running performance tests in practice.

\textbf{While the response time improves on larger function sizes, the stability of the tests is not affected significantly by the allocated memory.}
We note in \Cref{fig:response-time-example} that the \textit{ConfirmBooking} average response time is considerably faster when the function size is 512~MB or larger.
However, we do not observe any significant difference in the stability of the experiments (coefficient of variation) across different function sizes (Mann-Whitney U test $p > 0.05$).
This means that the amount of memory allocated for the function has an impact on its response time (expected), but exerts no significant influence on the stability of experiments.

\subsection{RQ3: Does the performance of serverless applications change over time?}

\begin{figure}[tb]
    \centering
    \includegraphics[width=\textwidth, trim=30 0 60 20, clip]{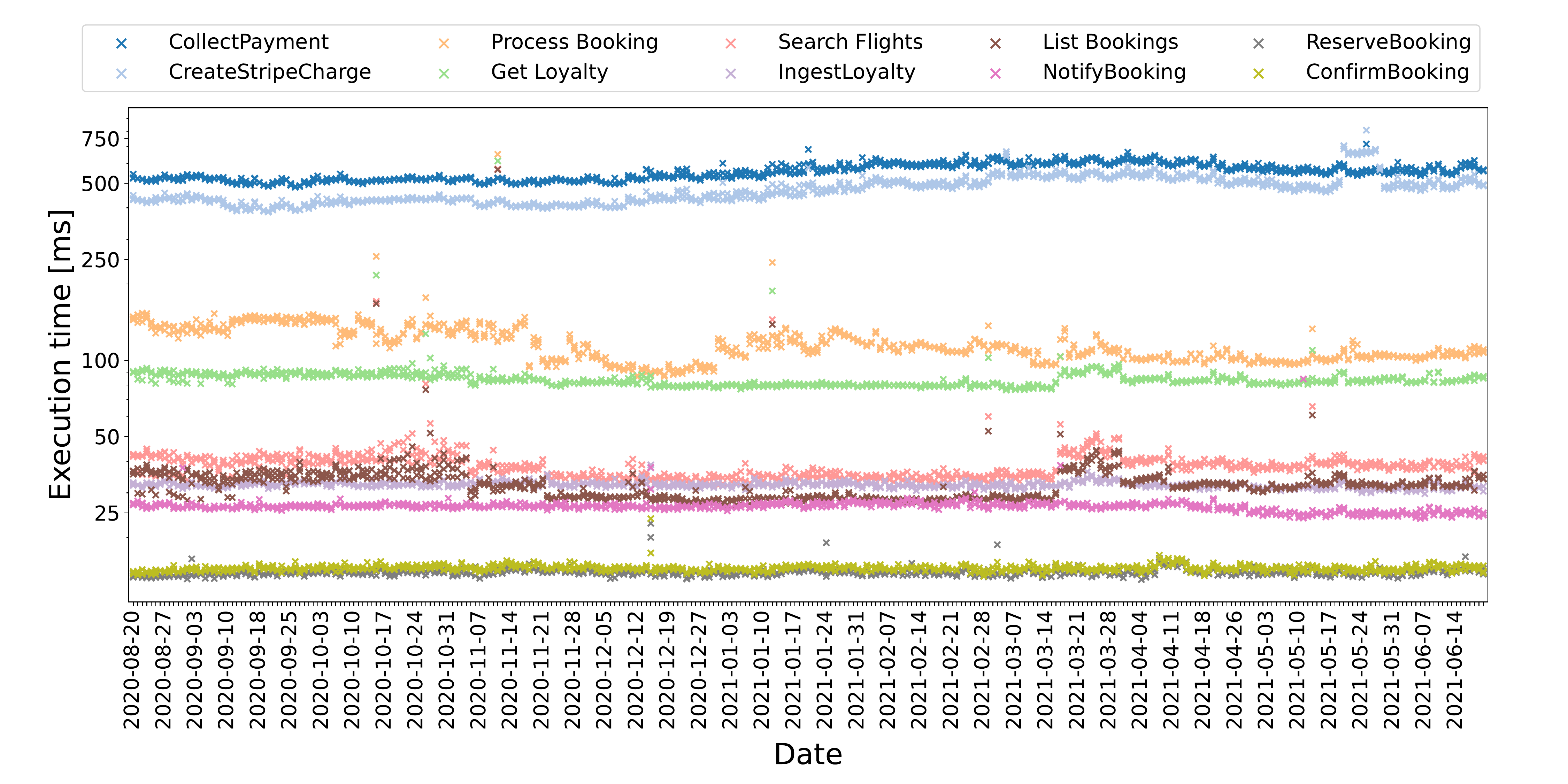}
    \caption{Mean response time for three daily performance measurements over a period of ten months.}
    \label{fig:longitudinal}
\end{figure}

\noindent \textbf{Motivation.} 
\label{sec:rq3}
RQ1 and RQ2 focus on the stability of performance tests conducted within the same time frame. However, the opaque nature of the underlying resource environments introduces an additional challenge: the underlying resource environment may change without notice.
This might result in both short-term performance fluctuations (e.g., due to changing load on the platform) or long-term performance changes (e.g., due to software/hardware changes). Therefore, in this RQ, we conduct a longitudinal study on the performance of our SUT, to investigate if we can detect short-term performance fluctuations and long-term performance changes.\\

\noindent \textbf{Approach.} 
We analyze the results of our longitudinal study (described in Section~\ref{sec:setup}), which consists of three measurement repetitions with 100 requests per second and 512 MB memory size every day for ten months. 
First, to determine if there are any significant changes in the distribution of the measurement results over time, we employ the change point detection approach from Daly et al.~\cite{daly2020Change}. To reduce the sensibility to short-term fluctuations, we use the median response time of the three daily measurements and configure the approach with $p=0$ and 100,000 permutations.
Second, upon visual inspection, it seemed that the variation between the three daily measurement repetitions was less than the overall variation between measurements. To investigate this, we conducted a Monte Carlo simulation that randomly picks 100,000 pairs of measurements that were conducted on the same day and 100,000 measurement pairs from different days. We calculated and compared the average variation between the sample pairs from the same day and from different days.
Finally, to investigate if the observed performance variation could be misinterpreted as a real performance change (regression), we conducted a second Monte Carlo simulation. We randomly select two sets of ten consecutive measurements that do not overlap and test for a significant difference between the pairs using the Mann–Whitney U test~\cite{mann1947}. For each detected significant difference, we calculate Cliff's Delta~\cite{Cliff} to quantify the effect size.
Similar to our first Monte Carlo simulation, we repeat this selection and comparison 100,000 times. 
Further implementation details are available in our replication package.\footnote{\url{https://github.com/ServerlessLoadTesting/ReplicationPackage}} \\

\begin{table}[tb]
    \centering
    \caption{Comparison of average performance variation between two measurements from either the same day or different days based on a Monte Carlo simulation.}
    \label{tab:same_vs_diff_day}
    \begin{tabular}{lrr}
    \toprule
    {Request class} &  Same-day Variation  & Overall Variation \\
    \midrule
ConfirmBooking & 2.1\% $\pm$ 2.6\% & 2.8\% $\pm$ 3.0\% \\
CreateStripeCharge & 2.2\% $\pm$ 2.1\% & 13.3\% $\pm$ 11.0\% \\
Get Loyalty & 3.0\% $\pm$ 20.0\% & 7.0\% $\pm$ 22.0\% \\
IngestLoyalty & 1.6\% $\pm$ 1.6\% & 2.9\% $\pm$ 2.4\% \\
List Bookings & 7.0\% $\pm$ 59.3\% & 16.2\% $\pm$ 65.3\% \\
NotifyBooking & 2.3\% $\pm$ 8.5\% & 4.3\% $\pm$ 8.4\% \\
Process Booking & 3.5\% $\pm$ 16.9\% & 17.2\% $\pm$ 20.4\% \\
ReserveBooking & 2.4\% $\pm$ 3.2\% & 3.2\% $\pm$ 3.8\% \\
CollectPayment & 1.9\% $\pm$ 2.0\% & 8.3\% $\pm$ 6.1\% \\
Search Flights & 7.1\% $\pm$ 50.1\% & 13.2\% $\pm$ 49.6\% \\
    \bottomrule
    \end{tabular}
\end{table}

\noindent \textbf{Findings.}
\textbf{There were short-term performance fluctuations during our longitudinal study, despite the fact that no changes were made to the application.} Figure~\ref{fig:longitudinal} presents the average response time of each API endpoint during the study periods. We can clearly observe fluctuations in performance. For example, the response time of the API \emph{Process Booking} has demonstrated large fluctuation after October 2020. Table~\ref{tab:same_vs_diff_day} compares the variation of performance between measurements from the same day and across different days (overall) using a Monte Carlo simulation. We find that in all of the API endpoints, the average variation between two random measurements is higher than the variation between measurements from the same day. For example, \emph{Process Booking} has an average variation of 17.2\% when considering all measurements, which is more than four times the average variation between measurements from the same day (3.5\%).

\begin{figure*}[t]
    \centering
    \begin{subfigure}{0.49\linewidth}
    \includegraphics[width=\textwidth]{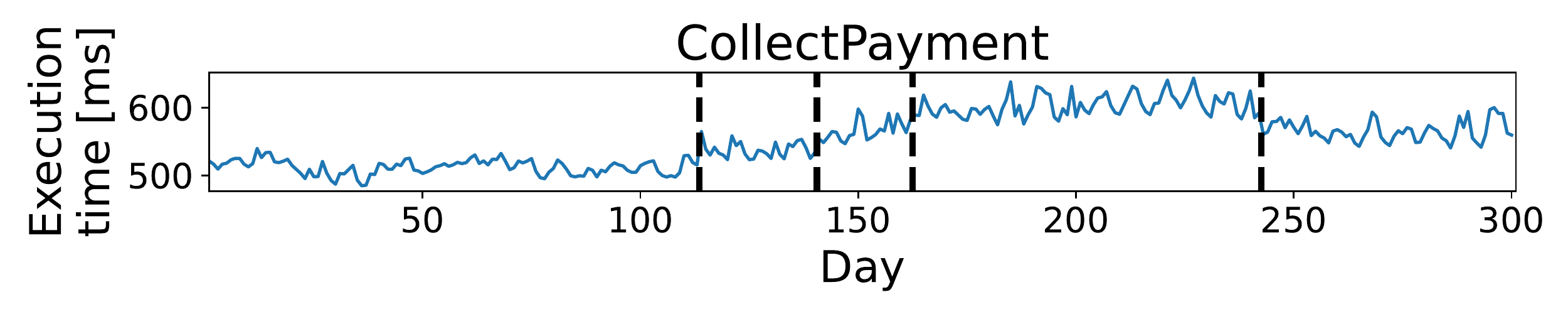}
    \end{subfigure}
    \begin{subfigure}{0.49\linewidth}
    \includegraphics[width=\textwidth]{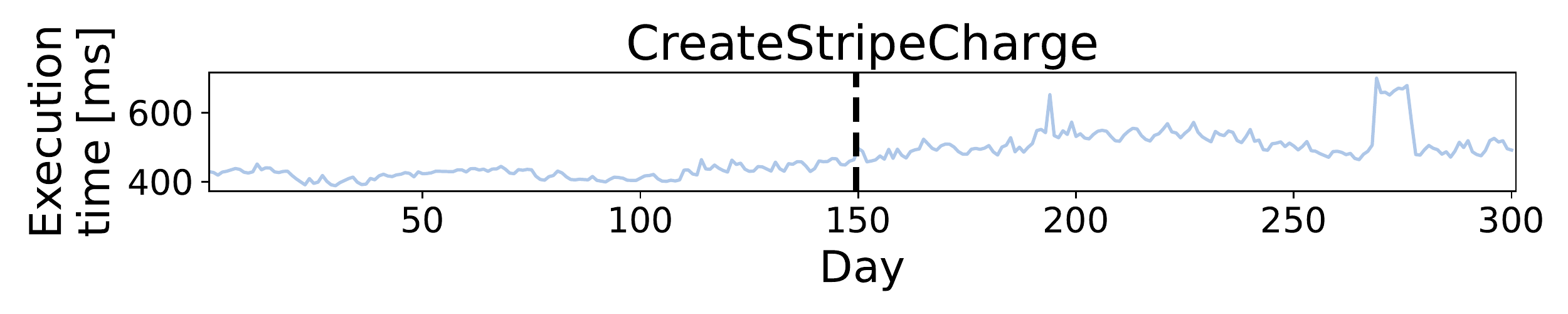}
    \end{subfigure}
    \begin{subfigure}{0.49\linewidth}
    \includegraphics[width=\textwidth]{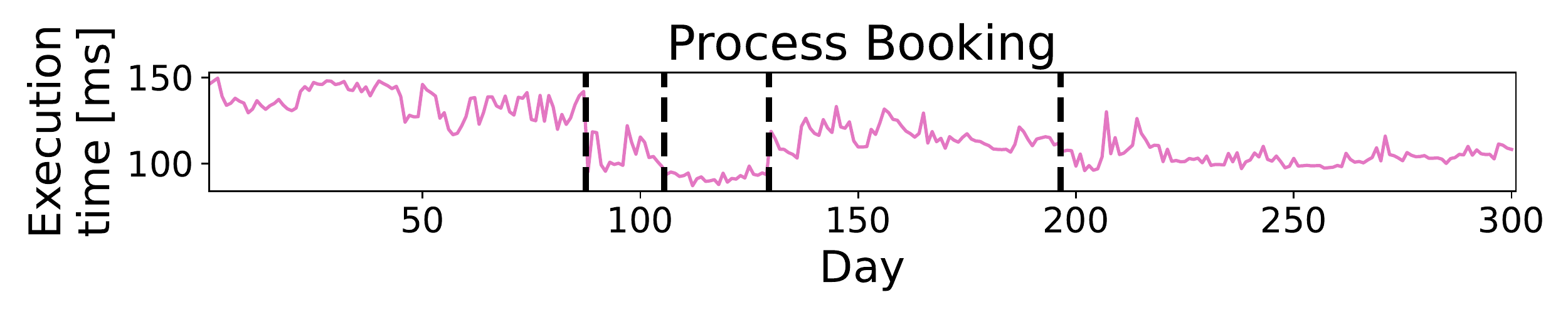}
    \end{subfigure}
    \begin{subfigure}{0.49\linewidth}
    \includegraphics[width=\textwidth]{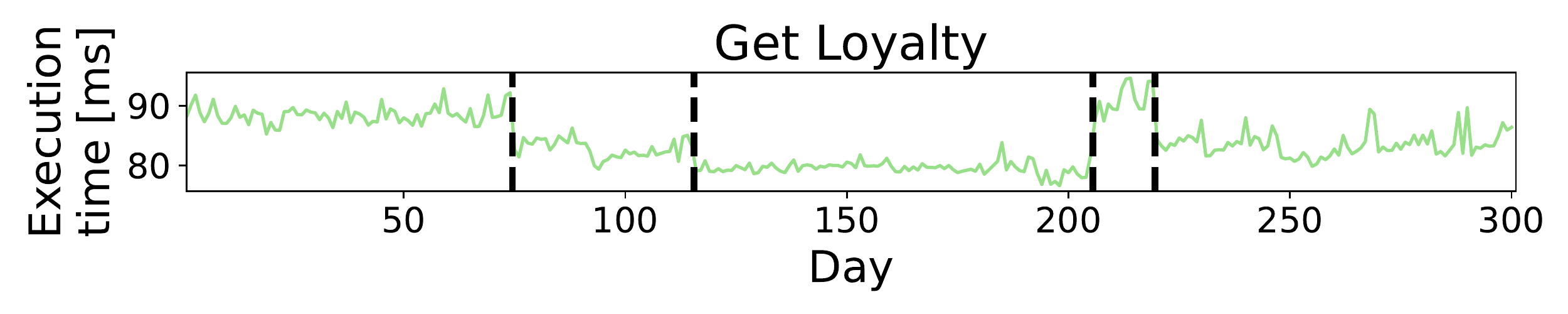}
    \end{subfigure}
    \begin{subfigure}{0.49\linewidth}
    \includegraphics[width=\textwidth]{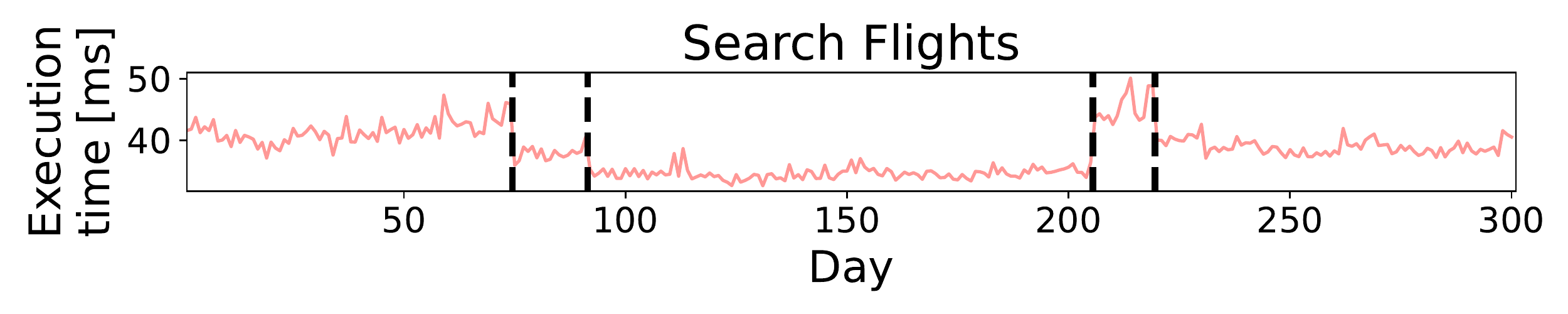}
    \end{subfigure}
    \begin{subfigure}{0.49\linewidth}
    \includegraphics[width=\textwidth]{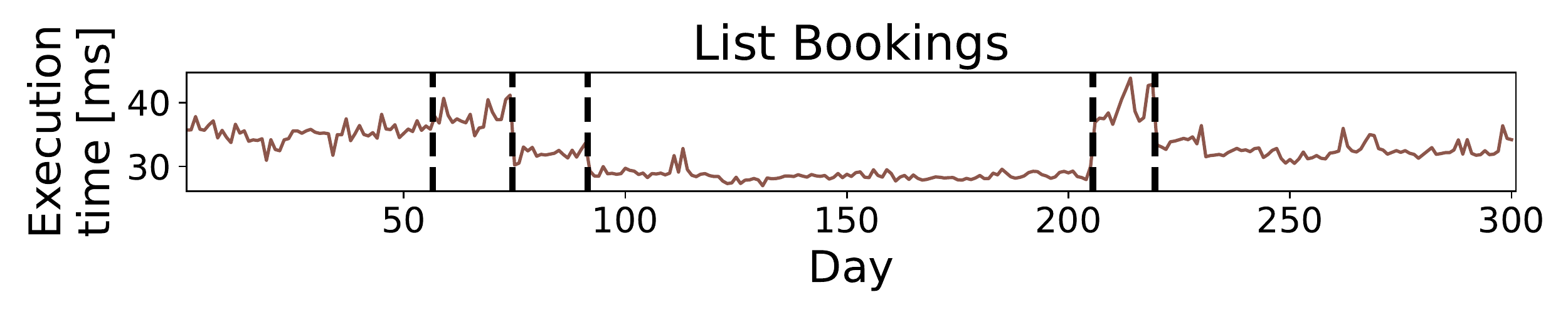}
    \end{subfigure}
    \begin{subfigure}{0.49\linewidth}
    \includegraphics[width=\textwidth]{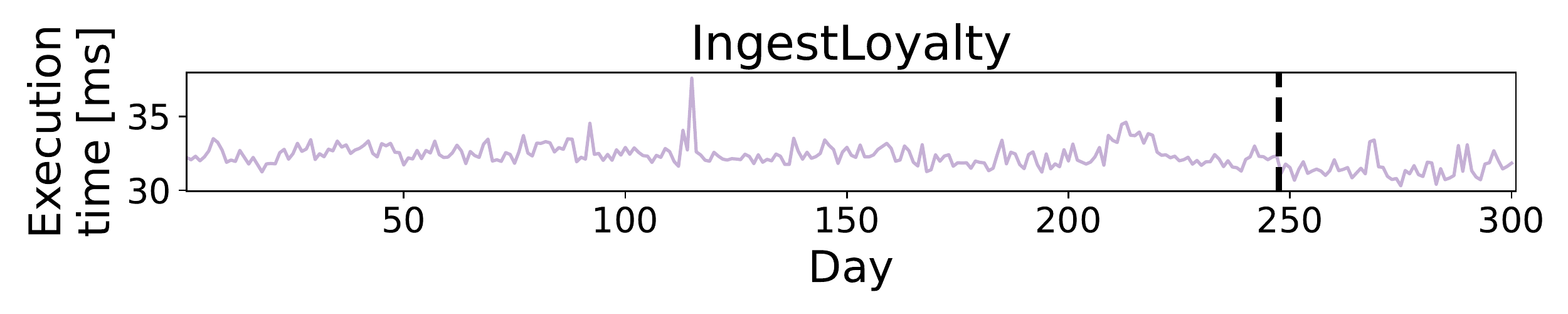}
    \end{subfigure}
    \begin{subfigure}{0.49\linewidth}
    \includegraphics[width=\textwidth]{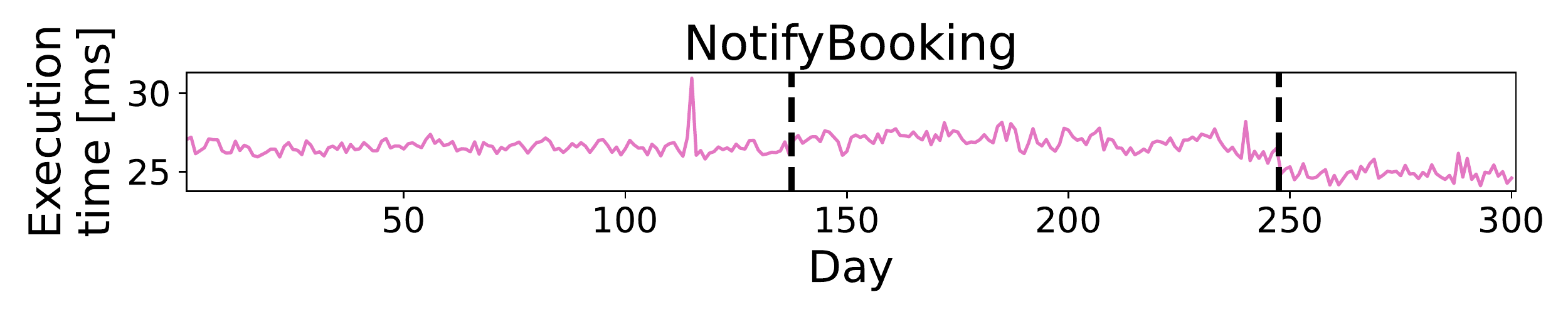}
    \end{subfigure}
    \begin{subfigure}{0.49\linewidth}
    \includegraphics[width=\textwidth]{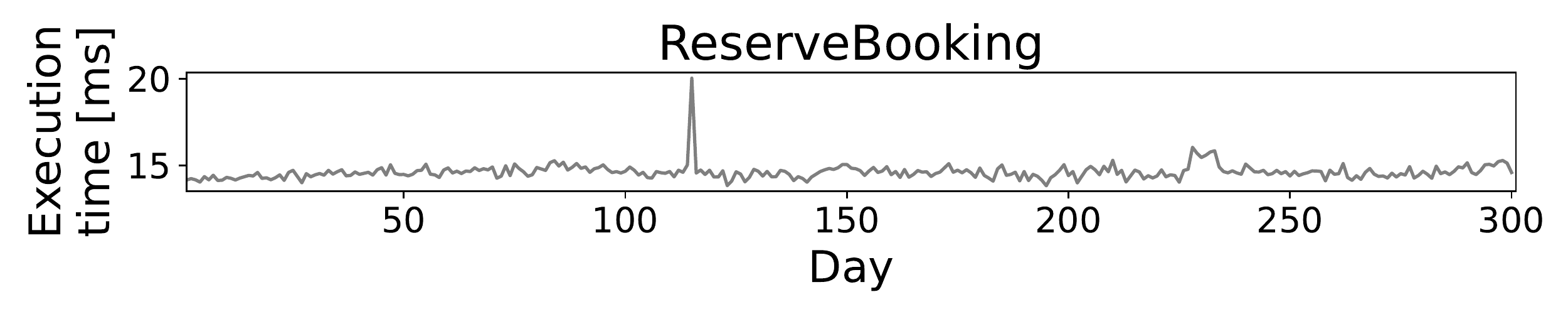}
    \end{subfigure}
    \begin{subfigure}{0.49\linewidth}
    \includegraphics[width=\textwidth]{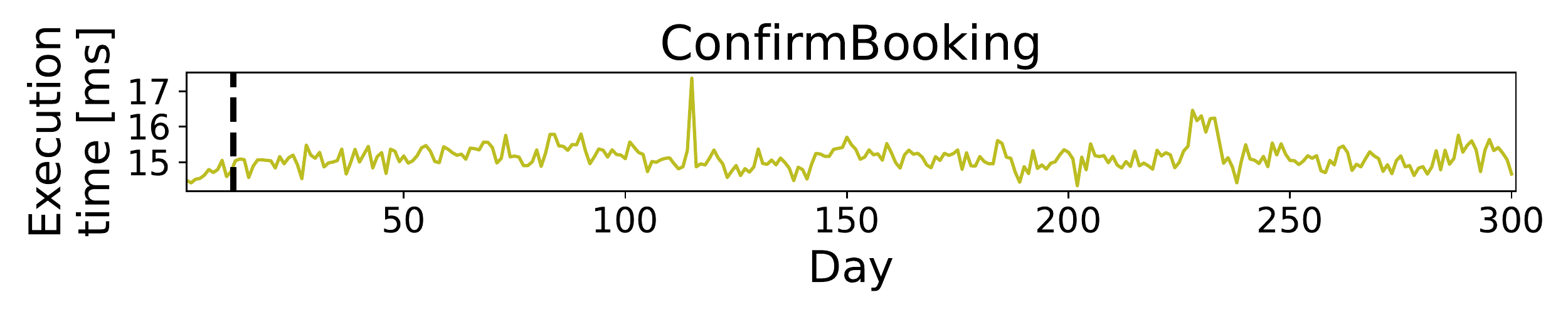}
    \end{subfigure}
    \caption{Detected change points for each workload class, note the different y-axis scales.}
    \label{fig:changepoints}
\end{figure*}

\textbf{We detect long-term performance changes during the observation period.} Figure~\ref{fig:changepoints} presents the detected long-term performance changes in the different APIs, according to the change point detection. Although some API endpoints have more change points than others, all of the API endpoints, except for \emph{ReserveBooking}, have gone through at least one change point (the change point in \emph{ConfirmBooking} might also be a false positive, as it is quite close to the experiment start). There exist as many as five change points during the observation period for an API endpoint. The impact of the performance change may be drastic. For example, the API endpoints \emph{Search Flights} and \emph{List Bookings} have similar performance changes where the response time is drastically reduced twice. On the other hand, the response time of some API endpoints, for example, \emph{CollectPayment}, increases in each change point, leading to a potential unexpected negative impact on the end-user experience. Finally, most of the change points for the different API endpoints do not appear at the same time, which may further increase the challenge of maintaining the performance of the serverless applications. 

\begin{table}[tb]
    \centering
    \caption{Percentage of at least negligible, small, or medium differences according to Mann-Whitney U test and Cliff's delta based on Monte Carlo simulation.}
    \label{tab:diffs-monte-carlo}
    \begin{tabular}{lrrrr}
    \toprule
    {Request class} &     Negligible+ &   Small+ &  Medium+ \\
    \midrule
    ConfirmBooking     &  54.0\% &  17.4\% &  6.8\% \\
    CreateStripeCharge &  11.0\% &  3.8\% &  1.7\% \\
    Get Loyalty        &  21.6\% &  7.1\% &  2.9\% \\
    IngestLoyalty      &  43.3\% &  14.0\% &  5.6\% \\
    List Bookings      &  19.2\% &  6.8\% &  3.1\% \\
    NotifyBooking      &  37.1\% &  11.9\% &  4.7\% \\
    Process Booking    &  14.0\% &  5.0\% &  2.3\% \\
    ReserveBooking     &  57.0\% &  18.5\% &  7.0\% \\
    CollectPayment     &  13.2\% &  4.4\% &  1.9\% \\
    Search Flights     &  24.3\% &  7.9\% &  3.3\% \\
    \bottomrule
    \end{tabular}
\end{table}

\textbf{The short-term performance fluctuations and long-term performance changes may have been considered as false performance regressions.} Table~\ref{tab:diffs-monte-carlo} shows the results of conducting the Mann-Whitney U test and measuring Cliff's delta between two groups of consecutive, non-overlapping samples based on our Monte Carlo simulation. We find that for four API endpoints, almost half of the comparisons have a statistically significant performance difference, even though the serverless application itself was \emph{identical} throughout the observation period. On the other hand, most of the differences have lower than medium effect sizes. In other words, the magnitude of the differences may be small and negligible, such that the impact on end users may not be drastic. However, there still exist cases whether large effect sizes are observed. Practitioners may need to be aware of such cases due to their large potential impact on end-user experience.

\section{Discussion}
\label{sec:discussion}
\begin{figure*}[t]
    \centering
    \includegraphics[width=\textwidth]{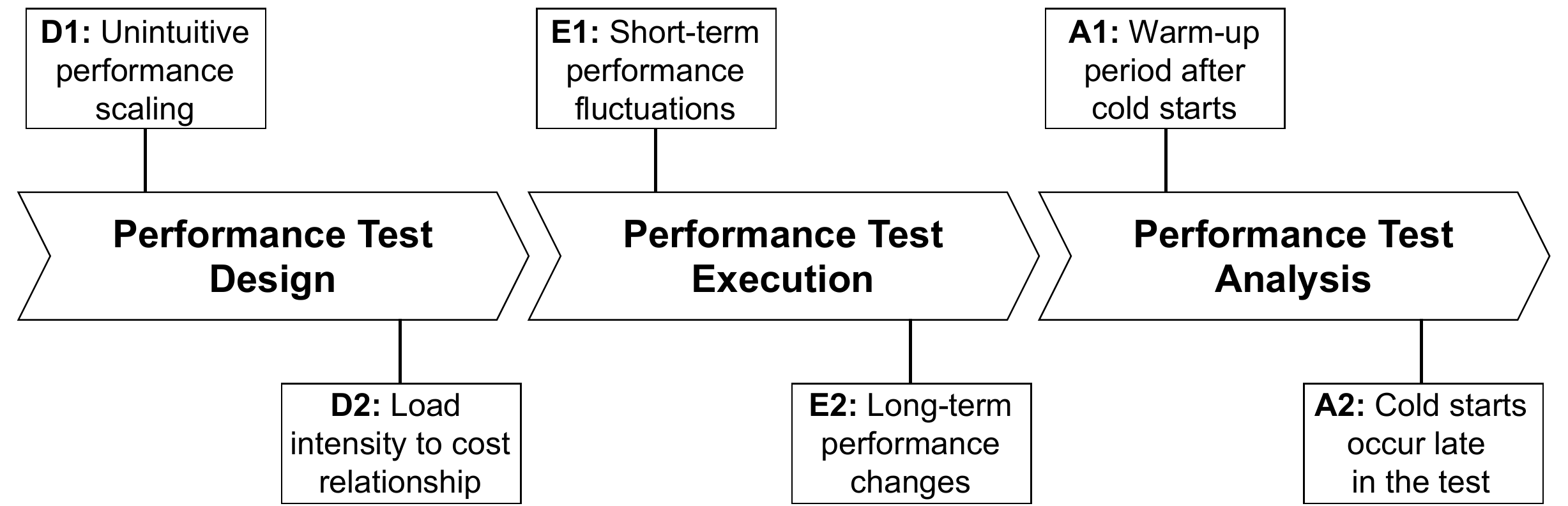}
    \caption{Properties of serverless that influence the different performance test stages.}
    \label{fig:results}
\end{figure*}

According to Jiang et al.~\cite{DBLP:journals/tse/JiangH15}, performance tests consist of three stages: (1) designing the test, (2) running the test, and (3) analyzing the test results. Based on the findings from our case study, we identified multiple properties of performance tests of serverless applications that practitioners should consider in each of these stages, as shown in Figure~\ref{fig:results}. 

\subsection{Design Phase}
During the design of a performance test, the key factors are the workload (which types of requests in which order), the load intensity (the number of requests), and the duration of the performance test.

\textbf{D1: Unintuitive performance scaling.} One of the key selling points of serverless platforms is their ability to seamlessly, and virtually infinitely scale with increasing traffic~\cite{eismann2020serverless}. Therefore, the classical approach of running performance tests at increasing load intensities, to see how much the performance deteriorates, becomes obsolete. We find in our experiments that the performance still differs at different load levels, however, and perhaps counterintuitively, the execution time decreases with increasing load. This property impacts how to plan performance tests. For example, a developer might run a performance test at 200 requests per second and find that the performance satisfies the SLA; however when the application is deployed and receives only 100 requests per second, it might violate the SLA. \textit{Therefore, developers need to consider that the worst performance is no longer observed at the highest load}. Depending on the use case, performance testing strategies could aim to: (a) quantify the expected performance by aiming to match the production load level, (b) understand how different load levels impact the performance by measuring a range of load intensities, or (c) aim to find the worst case performance with a search-based approach. 

\textbf{D2: Load intensity to cost relationship.} Traditionally, the cost of a performance test is independent of the load intensity and depends only on the number of deployed VMs and the duration of the experiment. For a serverless application, this relationship is inverted due to the pay-per-use pricing model of serverless. Due to this per-request pricing, the costs of a performance test has a linear relationship to the total number of requests in a performance test, for example, a performance test with 50 requests per second costs ten times as much as a performance test with 5 requests per second. This changes how developers should think about the costs of a performance test. Additionally, increasing the load intensity from five requests per second to 500 requests per second resulted in only a minor increase in result stability in our case study. Therefore, \textit{running more repetitions of a performance test at low load intensity instead of a single, large test could result in more stable results at the same cost}. However, further experiments in this direction are required to determine how much this increases the result stability.

\subsection{Execution Phase}
For the execution of a performance test, performance engineers need to decide when and how the test is executed. The technical implementation of a performance test is mostly unaffected by the switch to serverless applications, as most tooling for the performance testing of HTTP APIs (e.g., for microservice applications) can be reused. However, we find that there are two properties of serverless applications that influence the scheduling of performance tests.

\textbf{E1: Short-term performance fluctuations.} We find that the performance of serverless applications can suffer from short-term (daily) performance variation. While performance variation has also been observed for virtual machines~\cite{Iosup2011on, Leitner2016patterns}, we find that the variation between measurements conducted on different days is larger than for measurements conducted on the same day for serverless applications. Depending on the goal of a performance test, this has different implications. If the goal is to compare the performance of two alternatives (e.g., to answer the question if the performance of an application changed between two commits), then the measurements for both alternatives should be conducted on the same day. On the other hand, \textit{if the goal of a performance test is to quantify the performance of an application, the measurement repetitions should be spread across multiple days} as this will result in a more representative performance.

\textbf{E2: Long-term performance changes.} We detect a number of long-term performance changes that caused the performance of the application to permanently change in our case study, despite no changes being made to the application itself. We hypothesize that these performance changes are caused by updates to the software stack of the serverless platform; however, most serverless services do not offer any publicly available versioning that could be used to corroborate this. Unlike the short-term fluctuations, this issue can not be combated by running a larger number of measurement repetitions or by adopting robust measurement strategies such as multiple randomized interleaved trials~\cite{Abedi2017conducting}. When comparing two alternatives, they should be measured at the same time to minimize the chance of a long-term performance change occurring between the measurements, which is currently not necessarily the case, for example, for performance regression testing. \textit{Quantifying the performance of a serverless application is no longer a discrete task, but rather a continuous process, as the performance of a serverless application can change over time}.

\subsection{Analysis Phase}
In this phase, the monitoring data collected during the execution phase is analyzed to answer questions related to the performance of the SUT. A key aspect of this phase is the removal of the warm-up period to properly quantify the steady-state performance.

\textbf{A1: Warm-up period after cold starts.} The performance of a serverless application is generally separated into cold starts, which include initialization overheads, and warm starts, which are considered to have reached the steady-state phase and yield a more stable performance. We find that a performance test can still have a warm-up period even after excluding cold starts. A potential reason might be that, for example, caches of the underlying hardware still need to be filled before steady-state performance is reached.
This indicates that \textit{in the analysis of performance test results, the warm-up period still needs to be analyzed and excluded}. For our data, MSER-5, the current best practice to determine the warm-up period~\cite{ DBLP:conf/wsc/MahajanI04,White:00:MSER5}, was not applicable due to large outliers present in the data, a well-documented flaw of MSER-5~\cite{sandikcci2006analysis}. Therefore, future research should investigate suitable approaches for detecting the warm-up period of serverless applications.

\textbf{A2: Cold starts occur late in the test.} Another aspect about cold starts is that for a constant load, one could expect to find cold starts only during the warm-up period. In our experiments, we found that while the vast majority of cold starts occur during the warm-up period, some cold starts are scattered throughout the experiment. This might be, for example, due to worker instances getting recycled~\cite{lloyd2018serverless}. 
While these late cold starts did not significantly impact the mean execution time, they might impact more tail-sensitive measures such as the 99$^{th}$ percentile. Therefore, \textit{performance testers need to keep the possibility of late cold starts in mind while analyzing performance testing results}.

\section{Threats to Validity}
\label{sec:validity}
This section discusses threats to the construct, internal, and external validity of this study~\cite{Wohlin2012experimentation}. Construct validity examines the relation of the measurements to the proposed research questions. Internal validity examines the trustworthiness of the cause-and-effect relationship, that is, the existence of alternative explanations for findings, and external validity considers how well the results can be generalized.

\subsection{Construct Validity} 
In our experiments, we measured only the response time and function execution time; other metrics might show different effects. Out of the commonly used performance metrics, we did not consider CPU utilization and throughput. However, measuring the throughput is unusual for serverless applications due to their built-in scalability, and CPU utilization is currently not exposed by AWS. Further, we limited our experiments to performance tests with a constant load; performance tests with varying load might behave differently. Constant load is commonly used for performance tests, whereas varying load is more commonly used for load and stress testing. However, further research is required to understand the effects of performance tests under varying load.

\subsection{Internal Validity}
As the MSER-5 method for determining the duration of the warm-up period was not applicable to our data, we used a custom heuristic. It might be possible that this heuristic does not appropriately capture the length of the warm-up period. Based on a visual inspection of a large subset of the experiments, we found that the heuristic seems to capture the warm-up period well. Our replication package can be used to repeat this visual inspection. 

Another threat to the validity of our results is that performance experiments in the cloud can suffer from a high degree of uncertainty. To mitigate this threat, we followed recommended practices for conducting and reporting cloud experiments~\cite{Papadopoulos2019methodological} and used randomized multiple interleaved trials~\cite{Abedi2017conducting} to reduce measurement variability. Further, we provide a fully automated measurement harness that enables the replication of our measurements. For the longitudinal study, we perform three measurement repetitions each day at the same time to mitigate measurement variability, but we do not attempt to further control for performance variability as the study was intended to investigate the variability.

\subsection{External Validity}
Our case study used only a single SUT, which might limit the generalizability of our results. However, the serverless airline booking application is larger (uses more functions) than the average serverless application~\cite{eismann2020serverless, shahrad2020serverless}, so independent parts of the application could also be considered multiple applications. Further, most of the properties we measure are more dependent on the underlying cloud platform than the application itself. However, it is possible that a different application, such as a scientific computing application with long-running functions might behave differently. While our experiments were conducted on one application only, our methodology is applicable to any application. Another threat is that we conduct measurements on a single cloud platform. Although AWS is by far the most popular cloud provider for serverless applications, with 55\%-70\% of serverless applications running on AWS~\cite{eismann2020serverless, gtst, scs}, further research is required to determine if our findings are transferable between cloud providers.

\section{Replication Package}
\label{replication}
Performance measurements of public cloud environments are per definition only a snapshot of the performance at the time of measurement~\cite{Leitner2016patterns, Abedi2017conducting,Iosup2011on}. The performance properties can change whenever the cloud provider upgrades its hardware, switches to newer versions of the underlying operating system or virtualization technology, introduces new optimizations or features for the offered services, or changes any number of configuration parameters~\cite{Eismann2020microservices, Iosup2011on}. To increase our results' longevity, we provide a replication package that allows other researchers to replicate our findings and enables tracking if and how the reported performance properties evolve over time. This is in line with the recently proposed methodological principles for the reproducible performance evaluation of public clouds by Papadopoulos et al.~\cite{Papadopoulos2019methodological}.

Our replication package\footnote{\url{https://github.com/ServerlessLoadTesting/ReplicationPackage}} consists of two parts: (a)~the experiment harness used to run the performance measurements and (b)~the data and analysis scripts used in the presented analysis. We provide the experiment harness as a Docker container that replicates all measurements conducted in this study with a single CLI command from any Docker-capable machine. To simplify the reuse of this harness in other studies, experiments can be specified as JSON files, including measurement duration, load intensity, load pattern, measurement repetitions, and system configuration. The second part of our replication package is a CodeOcean capsule containing the collected measurement data and the scripts for the analysis presented in this paper. The CodeOcean capsule enables a one-click replication of our analysis either on the measurement data we collected or on new measurement data collected using our measurement harness.

\section{Conclusion}
\label{sec:conclusion}
Serverless applications delegate resource management tasks, such as deployment, resource allocation, or auto-scaling, to the cloud provider~\cite{kounev_et_al:DagRep.11.5.1:Ch5.1:ServerlessNotion,eismann2020review}, who bills users on a pay-per-use basis~\cite{baldini2017serverless, eyk2019Reference}. A common and powerful approach to manage system performance is the regular execution of performance tests; however, performance tests require that an identical resource environment is used for all tests, which cannot be guaranteed for a serverless application~\cite{Eismann2020microservices}. Therefore, we conducted an exploratory case study on the stability of performance tests of serverless applications, including a longitudinal study of daily measurements for ten months.

We find that there are serverless-specific changes and pitfalls to all performance test phases: design, execution, and analysis. In the design phase, the load intensity of the test directly correlates to cost, and reducing load intensity can decrease performance. In the execution phase, daily performance fluctuations and long-term performance changes impact the decision when performance tests should be scheduled. In the analysis phase, developers need to consider that there is still a warm-up period after removing all cold starts and that cold starts can occur late in a performance test under constant load.

\section*{Acknowledgments}
The work was conducted by the SPEC RG DevOps Performance Working Group.\footnote{\url{https://research.spec.org/devopswg}} This work was supported by the AWS Cloud Credits for Research program. The authors would like to thank Heitor Lessa for his support with the serverless airline booking application, as well as David Daly and Alexander Costas for their input on the change point detection.

\bibliographystyle{abbrvnat}
\bibliography{bibliography}
\end{document}